\documentclass{optica-article_test}

\journal{oe}

\articletype{Research Article}

\usepackage{caption}
\usepackage{subcaption}
\usepackage{lineno}
\usepackage{amsmath}
\usepackage{textgreek}
\usepackage{graphicx}
\usepackage{booktabs}
\usepackage{hyperref}
\hypersetup{
    colorlinks=true,
    linkcolor=blue,
    filecolor=magenta,      
    urlcolor=blue,
    }

\begin{document}

\title{A compact 20-pass thin-disk multipass amplifier stable against thermal lensing effects and delivering 330 mJ pulses with $\bf{M^2 < 1.17}$}

\author{Manuel Zeyen,\authormark{1} Lukas Affolter,\authormark{1} Marwan Abdou Ahmed,\authormark{2} Thomas Graf,\authormark{2} Oguzhan Kara,\authormark{1} Klaus Kirch,\authormark{1,3} Miroslaw Marszalek,\authormark{1} François Nez,\authormark{4} Ahmed Ouf,\authormark{5} Randolf Pohl,\authormark{5,6} Siddharth Rajamohanan,\authormark{5} Pauline Yzombard,\authormark{4} Karsten Schuhmann\authormark{1} and Aldo Antognini\authormark{1,3,*}}

\address{\authormark{1}Institute for Particle Physics and Astrophysics, ETH, 8093 Zurich, Switzerland.\\
\authormark{2}Institut für Strahlwerkzeuge, Universität Stuttgart, Pfaffenwaldring 43, 70569 Stuttgart, Deutschland.\\
\authormark{3}Laboratory for Particle Physics, Paul Scherrer Institute, 5232 Villigen, Switzerland.\\
\authormark{4}Laboratoire Kastler Brossel, Sorbonne Université, CNRS, ENS-Université PSL, Collège de France, 75252 Paris Cedex 05, France.\\
\authormark{5}QUANTUM, Institute of Physics, Johannes Gutenberg-Universität Mainz, 55099 Mainz, Germany.
\authormark{6}Excellence Cluster PRISMA+, Johannes Gutenberg-Universität Mainz, 55099 Mainz, Germany.\\
}

\email{\authormark{*}aldo.antognini@psi.ch}

\begin{abstract}

We report on an Yb:YAG thin-disk multipass amplifier delivering 50 ns long pulses at a central wavelength of 1030 nm with an energy of 330 mJ at a repetition rate of 100 Hz. The beam quality factor at the maximum energy was measured to be $\text{M}^2 = 1.17$. The small signal gain is 20, and the gain at 330 mJ was measured to be 6.9. The 20-pass amplifier is designed as a concatenation of stable resonator segments in which the beam is alternately Fourier transformed and relay-imaged back to the disk by a 4f-imaging optical scheme stage. The Fourier transform propagation makes the output beam robust against spherical phase front distortions, while the 4f-stage is used to compensate the thermal lens of the thin-disk and to reduce the footprint of the amplifier.
\end{abstract}

\section{Introduction}
Multipass amplifiers are versatile tools for the generation of high-power and high-energy laser beams. Systems delivering kWs of average power have been demonstrated at all pulse lengths from continuous wave down to femtoseconds. In particular, thin-disk multipass amplifiers have shown their great potential in combining high power/energy output with diffraction limited beam quality and high flexibility regarding the input pulse length \cite{Zapata:15, Herkommer:20, Nagel:21}.

Most thin-disk multipass amplifier designs are based on three main categories: relay-imaging (4f-imaging) \cite{siebold_high-energy_2012, ochi_effective_2017, perevezentsev_matrix_2017, papadopoulos_high_2015, Zwilich:20, druon_comparison_2020}, quasi-collimated propagation (lens guide) \cite{nagel_thin_2019, negel_ultrafast_2015, schulz_pulsed_2012, dietz_ultrafast_2020, hornung_54_2016} or resonator-based optical Fourier transform (FT) propagations \cite{antognini_thin-disk_2009, Schuhmann_thin-disk_2015, chyla_generation_2018}. In 4f-based designs, the beam is imaged from active medium (AM) to active medium so that its size remains constant for each pass (if soft aperture effects are neglected). These systems are relatively compact and input beams with a wide variety of beam parameters can be propagated. However, phase front distortions introduced by thermal effects in the AM add up with each pass over the AM, limiting the achievable beam quality and stability of such an amplifier. In the quasi-collimated design the beam propagates almost freely in the amplifier. Only weakly focusing elements are used to compensate the beam divergence and ensure the same size of the beam on each pass over the AM. This type of multipass amplifier is relatively convenient to set up and has the additional advantage that the propagating beam stays large along the whole propagation path, which makes the quasi-collimated scheme an excellent choice for the highest power applications with excellent beam quality \cite{Loescher:22}. While this design is less sensitive to phase front distortions at the AM than the 4f-design, a third design based on FT propagations provides the highest stability against both spherical and aspherical phase front distortions \cite{schuhmann_multipass_2018,zeyenthin}.

The FT design can be understood as an unfolded FT resonator of a corresponding regenerative amplifier. The amplifier thus inherits the stability of the FT resonator w.r.t. thermal lensing so that FT-based multipass amplifiers have the potential to deliver near diffraction limited output beams at kW (Joule) level output power (energy). However, a resonator-based multipass amplifier typically requires careful setup and mode matching of the input beam to the resonator mode, and a rather large mirror array.

Here we report on the development of a multipass amplifier that combines the advantages of the 4f-based imaging approach with the stability of the FT-resonator design. This hybrid amplifier is part of  the laser system of the HyperMu experiment, in which the CREMA (Charge Radius Experiments with Muonic Atoms) collaboration aims at measuring the ground-state hyperfine splitting of muonic hydrogen~\cite{amaro_laser_2022,nuber2022diffusion}. In this system a single-frequency Yb:YAG thin-disk laser (TDL), composed of an oscillator followed by the here presented amplifier, is used for pumping mid-IR optical parametric oscillators (OPOs) and optical parametric amplifiers (OPAs). The input laser pulses to the amplifier are provided by an injection seeded TDL oscillator which is frequency stabilized via the Pound-Drever-Hall method \cite{Zeyen2023, Zeyen2023a}. The TDL oscillator delivers single-frequency pulses of up to 50\,mJ energy and a pulse length between 50-110\,ns with nearly diffraction limited beam quality. Ultimately, the goal is to develop an amplifier which delivers nanoseconds pulses with 500\,mJ of pulse energy and an excellent beam quality factor together with a very high long-term energy/power and pointing stability.

In the following, Section~\ref{sec:Concept} introduces the multipass amplifier concept, and Section~\ref{sec:Layout} further elaborates on the optical layout. In Section~\ref{sec:Realization}, we describe the experimental realization of our hybrid multipass amplifier and discusses the advantages of including a 4f-stage in the propagation scheme. Finally the experimental results are presented in Section~\ref{sec:Results}.
\section{Amplifier concept} \label{sec:Concept}
The amplifier architecture includes a double 4f-imaging stage and an optical Fourier transform~(FT) segment, so that the propagation follows the sequence
\begin{equation}
4f - \text{ disk } - \text{FT} - \text{ disk } - 4f - 4f - \text{ disk } - \text{FT} - \text{ disk } \cdots,
\label{eq:prop}
\end{equation}
where the disk forms the beginning and the end of the FT propagation. Neglecting thermal lensing and aperture effects in the disk, the ABCD matrix of the basic double-pass segment $4f - \text{disk} - \text{FT} - \text{disk} - 4f$ is conveniently given by
\begin{gather}
M = M_\text{4f} M_\text{FT} M_\text{4f} = \begin{pmatrix}
-1 & 0 \\
0 & -1
\end{pmatrix}
\begin{pmatrix}
0 & F \\
-\frac{1}{F} & 0
\end{pmatrix}
\begin{pmatrix}
-1 & 0 \\
0 & -1
\end{pmatrix}
=  \begin{pmatrix}
0 & F \\
-\frac{1}{F} & 0
\end{pmatrix},
\label{eq:M}
\end{gather}
where $F$ is the so-called Fourier parameter \cite{schuhmann_multipass_2018, zeyenthin}. Only a collimated input beam with a radius~(1/$\text{e}^2$) of 
\begin{equation}
w_0=\sqrt{\frac{\lambda F}{\pi}}
\label{eq:w0}
\end{equation}
is reproduced by the ABCD matrix \eqref{eq:M} of this double-pass segment. In general, this makes resonator-based multipass amplifiers less flexible compared to 4f-based designs.

\begin{figure}
\centering
\includegraphics[width=\textwidth]{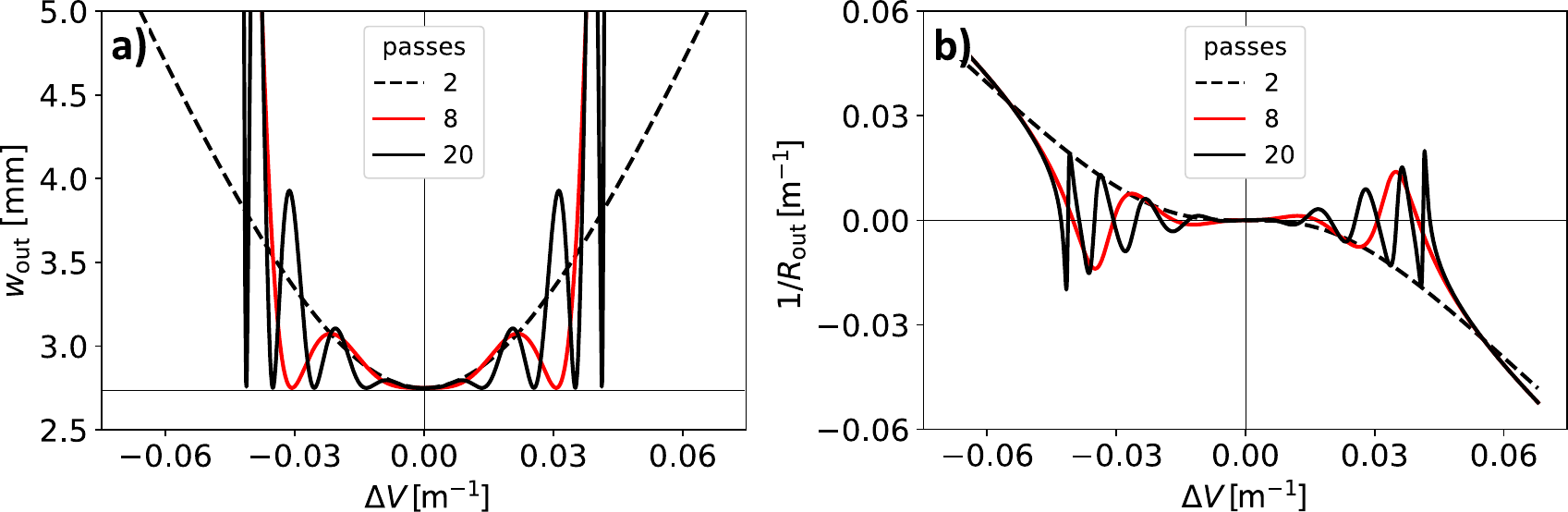}
\caption{Variation of the output beam size \textbf{a)} and phase front curvature \textbf{b)} against changes $\Delta V$ in the focal power of the disk for 2-, 8- and 20-passes in our hybrid-amplifier. The Fourier propagation of the amplifier was setup to accomodate an input beam of waist $w_0 = 2.75$\,mm, corresponding to $F=23.1$\,m. \label{fig:params_N}}
\end{figure}

\begin{figure}
\centering
\includegraphics[width=\textwidth]{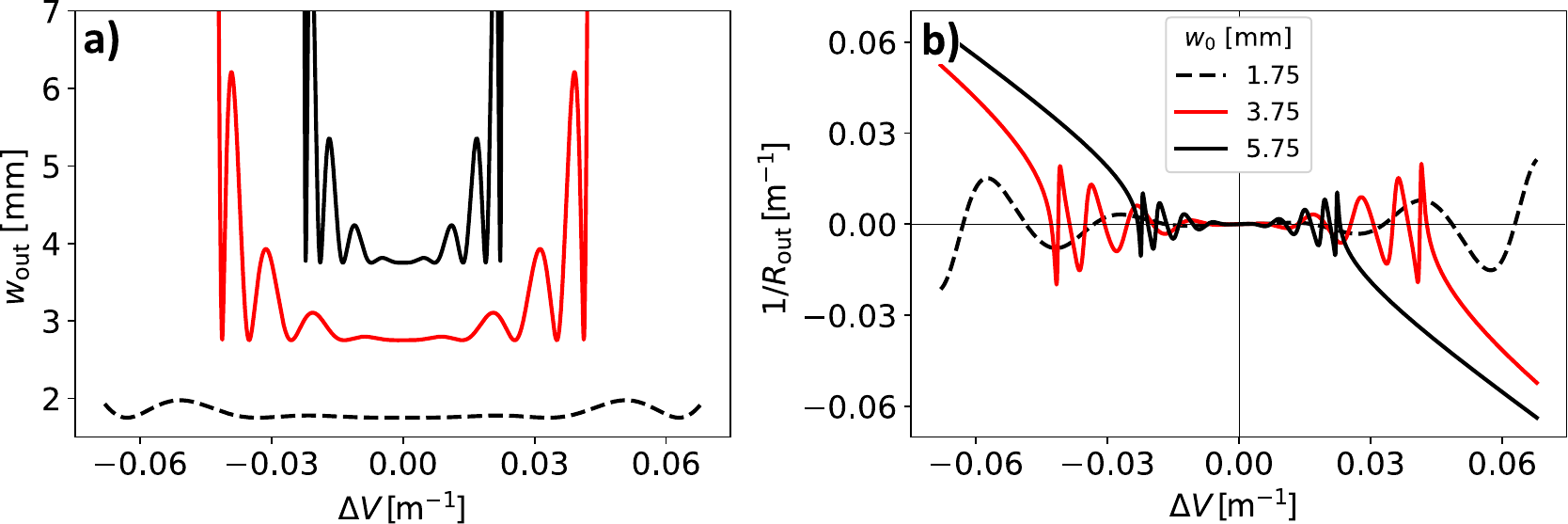}
\caption{Variation of the output beam size \textbf{a)} and phase front curvature \textbf{b)} against changes $\Delta V$ in the focal power of the disk for various beam sizes in a 20-pass hybrid-amplifier with $w_0 =$~1.75, 2.75 and 3.75\,mm, corresponding to F = 9.3, 23.1 and 42.9\,m, respectively. \label{fig:params_w0}}
\end{figure}

However, the advantage of the FT-resonator based approach becomes apparent when the disk's focal power (i.e., inverse focal length) changes by a small amount $\Delta V$, e.g. due to thermal lensing or variations of is curvature in the manufacturing process. Indeed, the FT propagation stabilizes the output beam parameters (size and phase front curvature) against $\Delta V$. After a single round-trip through a double-pass segment, the sensitivity of the beam parameters to $\Delta V$ is small, and given by \cite{schuhmann_multipass_2018}
\begin{equation}
\frac{w_\text{out}(\Delta V)}{w_0} = 1 + \frac{1}{2}F^2 \Delta V^2 - \frac{1}{8} F^4 \Delta V^4 +\ldots,
\end{equation}
\begin{equation}
\frac{1}{R_\text{out}(\Delta V)} = -F^2 \Delta V^3 + F^4 \Delta V^5 - \ldots,
\end{equation}
where $F$ is given by Eq.~\eqref{eq:w0}. For a 20-pass amplifier, as used in this study, the sensitivity of the beam parameters is
\begin{equation}
\frac{w_\text{20-pass}(\Delta V)}{w_0} = 1 + 50 F^4 \Delta V^4 + \ldots,
\label{eq:w_20-pass}
\end{equation}
\begin{equation}
\frac{1}{R_\text{20-pass}(\Delta V)} = 10 F^2 \Delta V^3 +\ldots\,.
\label{eq:C_20-pass}
\end{equation}
As illustrated in Fig.~\ref{fig:params_N}, the beam parameters change only marginally for small $\Delta V$ (in the stability zone), and the width of this stability zone is independent of $N$ contrary to the 4f-based design~\cite{antognini_thin-disk_2009,schuhmann_multipass_2018}. As visible in Fig.~\ref{fig:params_w0} the width of the stability zone strongly depends on $w_0$ which is a consequence of the design of the FT multipass amplifier as a resonator. Nevertheless, the stability zone is still noticeable in contrast to purely 4f-based designs \cite{schuhmann_thin-disk_2017, zeyenthin}. The stability against small changes of the focal power of the disk and the suppression of higher order aberrations by the FT not only make long-term operation of the amplifier robust but also simplify commissioning of the system (the amplifier can be initially setup at zero pump load and the beam shape always remains Gaussian).

\section{Optical layout} \label{sec:Layout}
\begin{figure}[h!]
\centering\includegraphics[width=0.75\linewidth]{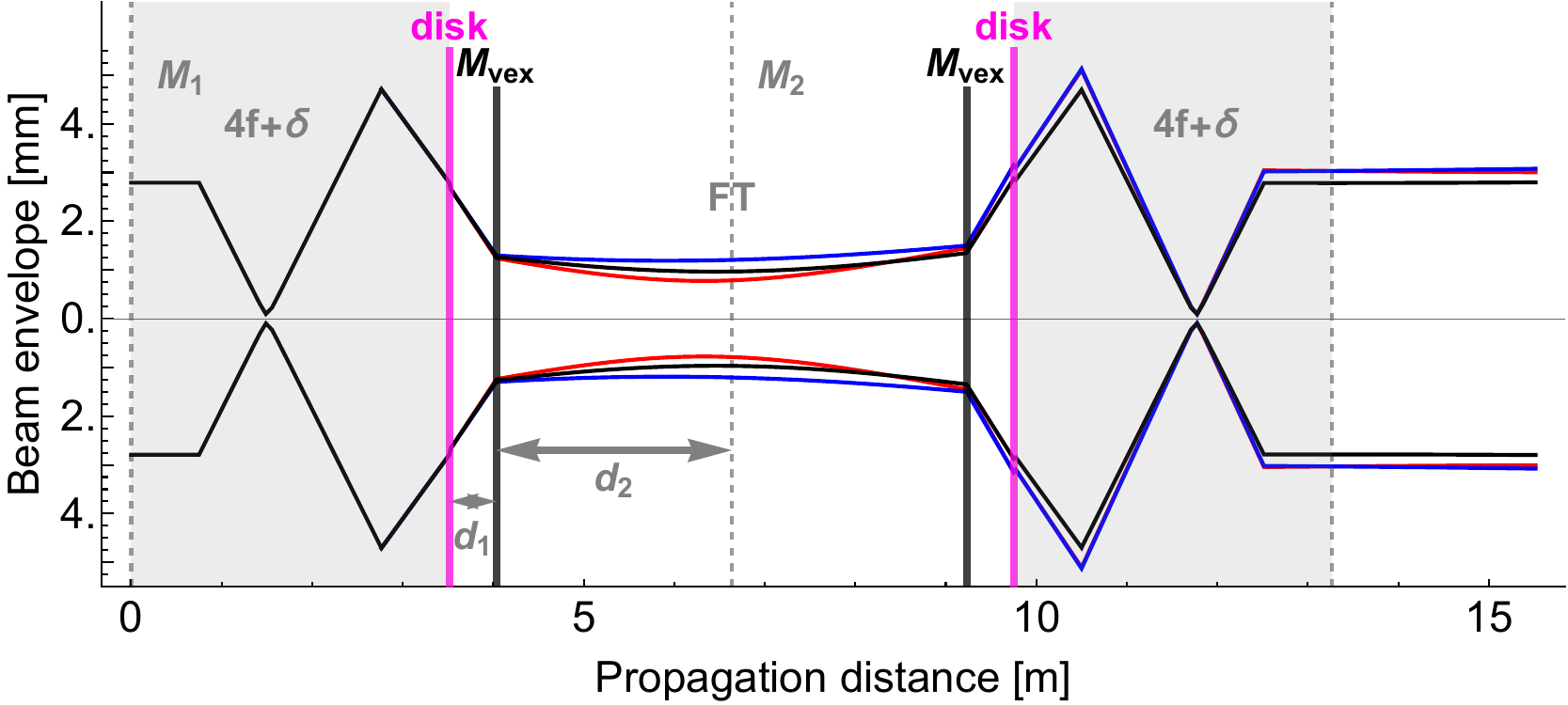}
\caption{Beam envelope (1/$\text{e}^2$ intensity-radius) in the basic double-pass segment of our hybrid amplifier with a detuned 4f-stage. The input beam has a waist of 2.75\,mm. The location of the disk is shown in magenta. The black line indicates the beam size for a disk with nominal focal power, i.e., $\Delta V = 0$. The red and blue lines are propagations for $\Delta V = \pm 0.01\text{ m}^{-1}$.}
\label{fig:BeamSize_basic_segment}
\end{figure}

In practice, it is crucial to shorten the FT propagation by realizing it as a Galilean telescope \cite{schuhmann_multipass_2018}. Such a telescope can be realized by a concave-shaped disk and a convex-shaped mirror, with the FT being performed over a round-trip through the telescope, i.e., from just before the disk back to just after the disk. The layout of the telescope is then calculated by matching the Galilean telescope to the ABCD matrix given in Eq.~\eqref{eq:M}. A realization of such an amplifier is shown in Fig.~\ref{fig:BeamSize_basic_segment} where the separation $d_1$ between the disk and the convex mirror and the separation $d_2$ between the convex mirror and the end mirror are given by \cite{zeyenthin}
\begin{align}
d_1 & =f_\text{vex} + f_\text{D}\frac{F}{F - f_\text{D}}, \label{eq:d1}\\
d_2 & =f_\text{vex} +\frac{1}{2}\left( \frac{f_\text{vex}}{f_\text{D}}\right)^2 F - \frac{f_\text{vex}^2}{2F},
\label{eq:d2}
\end{align}
where $f_\text{D}$ is the disk's focal length, and $f_\text{vex}$ is the focal length of the convex mirror. The beam waist $w_2$ on the end-mirror $\text{M}_2$ is given by
\begin{equation}
w_2 = \frac{\sqrt{2}}{2} w_0 f_\text{vex} \left(\frac{1}{F} -\frac{1}{f_\text{D}}\right),
\label{eq:w2}
\end{equation}
which serves to evaluate potential issues related to laser-induced damage.
While it is typically advantageous to realise the Galilean telescope in the FT branch with a curved disk, we had to use flat disks. Indeed, to maximize the extractable energy from the disk, i.e., to minimize amplified spontaneous emission effects, we used a rather thick and low-doped disk (600~\textmu m, 2.3 at.\%) \cite{zeyenthin}. It turned out to be problematic to contact such disks on curved diamond heat-spreaders so we contacted the disk on a flat diamond. The resulting focal length of the disk was $f_\text{D}\approx 10-15$\,m, which is too large for a standard Galilean telescope setup. However, if the 4f-stage is detuned, i.e., the separation between the focusing elements is changed from $2f$ to $2f+\delta$, an effective lens is produced at the disk. Indeed, the ABCD matrix of such a detuned 4f-stage is given by
\begin{gather}
M_\text{4f}(\delta) = \begin{pmatrix}
1 & f \\
0 & 1
\end{pmatrix}
\begin{pmatrix}
1 & 0 \\
-\frac{1}{f} & 1
\end{pmatrix}
\begin{pmatrix}
1 & 2f+\delta \\
0 & 1
\end{pmatrix}
\begin{pmatrix}
1 & 0 \\
-\frac{1}{f} & 1
\end{pmatrix}
\begin{pmatrix}
1 & f \\
0 & 1
\end{pmatrix}
=  \begin{pmatrix}
-1 & 0 \\
\frac{\delta}{f^2} & -1
\end{pmatrix},
\label{eq:M_4f_delta}
\end{gather}
which is equivalent to a 4f-stage followed by a thin lens of focal length $f' = f^2/\delta$. Placing the disk directly after the detuned 4f-stage thus allows to tune the disk's effective focal length. With a detuned 4f-stage, Eq.~\eqref{eq:d1} and Eq.~\eqref{eq:d2} are still valid when replacing 
\begin{equation}
\frac{1}{f_\text{D}} \rightarrow \frac{1}{f_\text{eff}} = \frac{1}{f_\text{D}} + \frac{1}{f'}.
\label{eq:f_eff}
\end{equation}
The output beam parameters are again stable against variations of $\Delta V$ since the round-trip ABCD matrix still has the form given by \eqref{eq:M}.

During operation, the effective lens introduced by the 4f-stage can also be employed to compensate for deviations of the focal power of the disk from its design value (e.g. due to thermal lensing under high pump load or manufacturing uncertainties) \cite{zeyenthin, neuhaus_passively_2009}. The detuning $\delta = \delta_\text{D} + \delta_\text{V}$ can be split into two parts: a fixed part
\begin{equation}
\delta_\text{D} = f^2 \left( \frac{1}{f'} -  \frac{1}{f_\text{D}}\right),
\end{equation}
used to set $f_\text{eff}$ to realize the nominal design, and a variable part
\begin{equation}
\delta_\text{V} = -f^2 \Delta V
\label{eq:det_4f}
\end{equation}
used to compensate for uncertain $\Delta V$. Operation in the center of the stability zone of the amplifier can be achieved by fine tuning $\delta$, resulting in optimal stability of the output beam parameters against thermal lensing.

Figure~\ref{fig:BeamSize_basic_segment} shows the calculated beam size in the basic double-pass segment of our hybrid amplifier. While the black curves indicate the beam size for a disk with nominal focal power $1/f_\text{D} \approx 0.01\text{ m}^{-1}$ (i.e., about 10\,m focal length), the red and blue curves illustrate the situation where the disk's focal power was changed by $\Delta V = \pm 0.02\text{ m}^{-1}$, respectively. The 4f-stage is realized with two concave mirrors of focal length $f=0.75$\,m. The separation between the two concave mirrors was increased by $\delta = 0.515$\,m to yield $f_\text{eff}\approx 1.5$\,m. The FT propagation was designed to keep the beam radius (1/$\text{e}^2$) on the disk at $w_0 = 2.75$\,mm while preserving a large enough waist $w_2 \approx 0.8$\,mm on the end-mirror $\text{M}_2$. The convex mirror used in the Galilean telescope of the FT branch has focal length $f_\text{vex} = -0.5$\,m. Inserting these numbers into Eq.~\eqref{eq:d1} and Eq.~\eqref{eq:d2} yields $d_1 = 0.52$\,m and $d_2=2.6$\,m, respectively.

\begin{figure}[h!]
\centering\includegraphics[width=\linewidth]{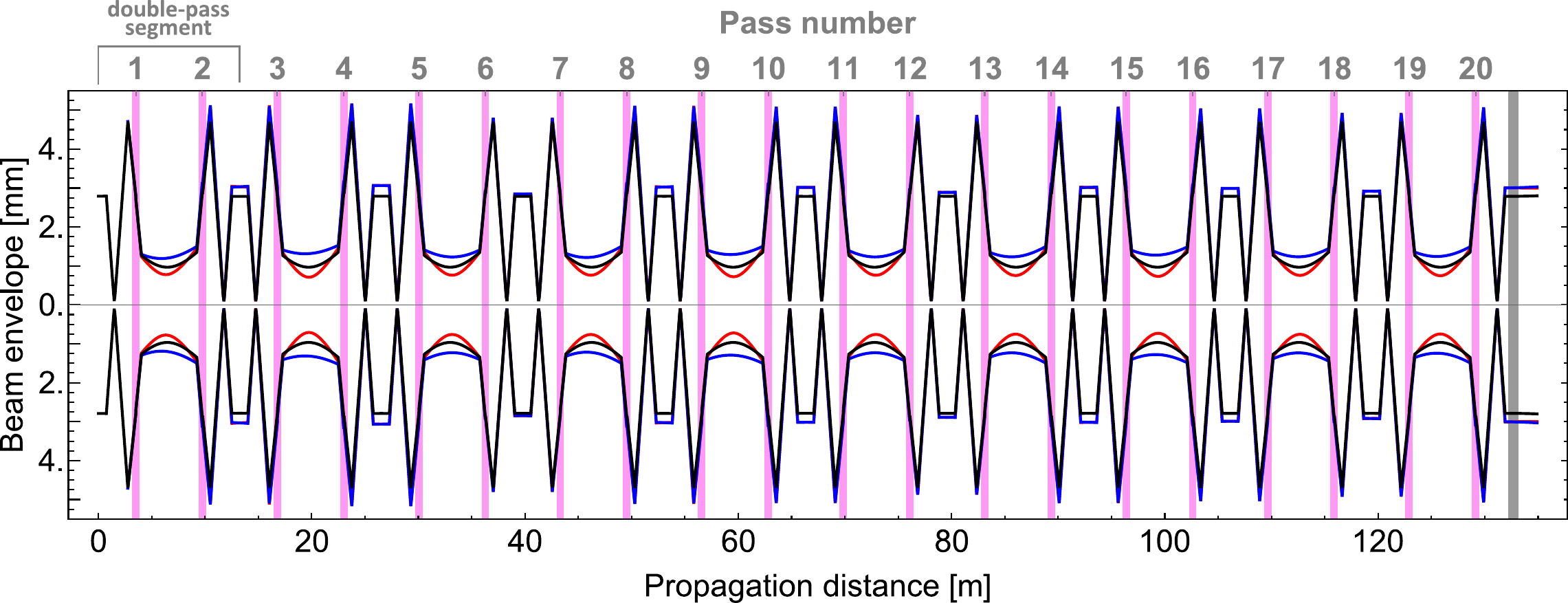}
\caption{Beam envelope in our 20-pass hybrid amplifier. The location of the disk is shown in magenta. The black line indicates the beam size for a disk with nominal focal power, i.e., $\Delta V = 0$. The red and blue lines are propagations for $\Delta V = \pm 0.01\text{ m}^{-1}$.}
\label{fig:BeamSize_20pass}
\end{figure}

Figure~\ref{fig:BeamSize_20pass} shows the calculated beam size along the whole 20-pass amplifier. The comparison between the various curves (defined as in Fig.~\ref{fig:BeamSize_basic_segment}) demonstrates that the beam size at the disk as well as the output beam size and its divergence are insensitive to changes of the disk's focal power $\Delta V$. This stability is provided by the FT part of the propagation in our hybrid multipass amplifier.

\section{Experimental realization} \label{sec:Realization}

\begin{figure}[h!]
\centering\includegraphics[width=\linewidth]{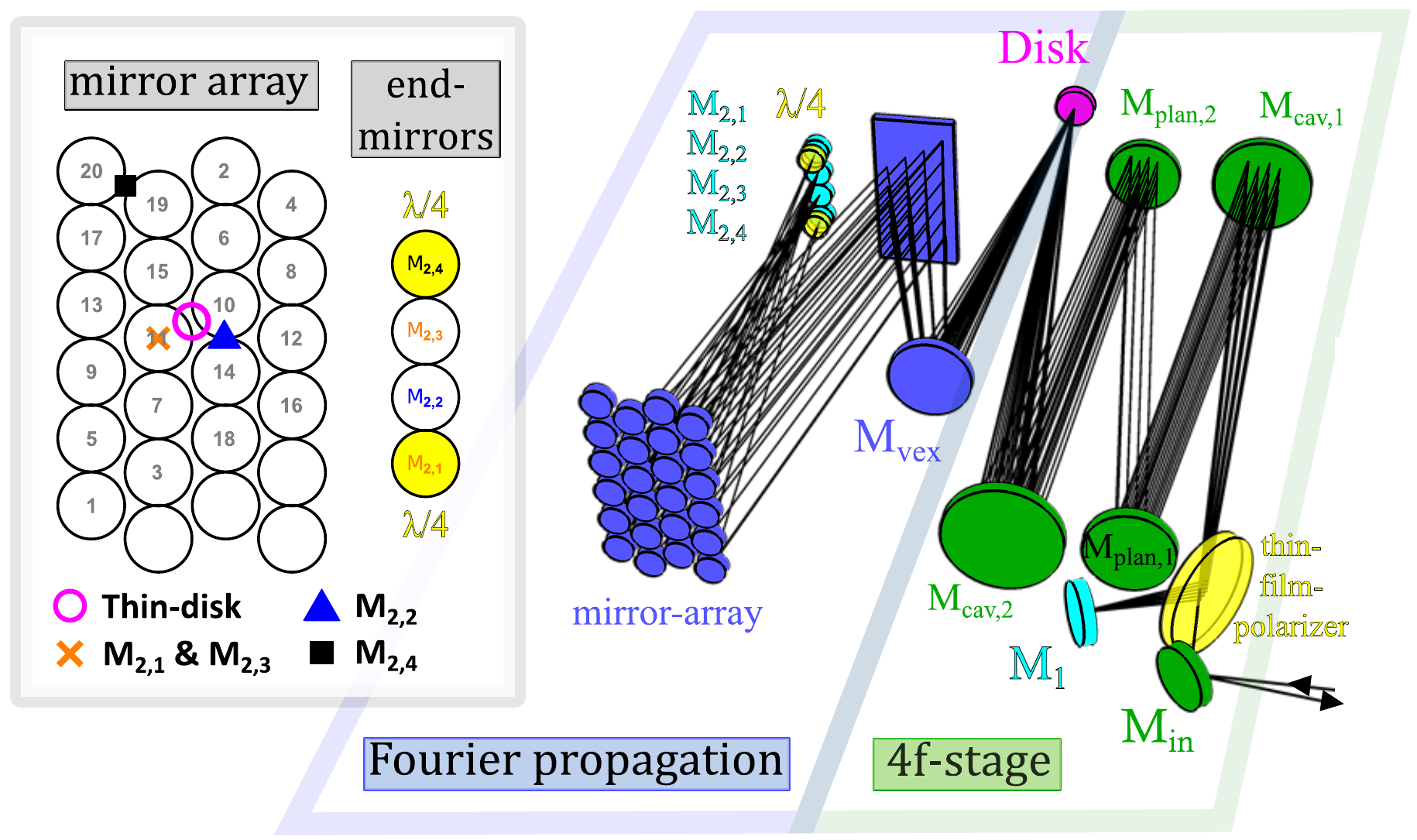}
\caption{Schematic of the hybrid 20-pass amplifier. Color coding: Disk -- magenta, 4f-stage -- green, Fourier propagation -- blue, end-mirrors -- cyan, thin film polarizer and wave plates -- yellow. Inset: Mirror array with the mirrors numbered in order of use. The symbols indicate where the symmetry axis of the corresponding optical element intersects the plane of the mirror array. All four locations act as point reflectors leading to the indicated propagation sequence. The yellow shaded mirrors of the end-mirror array have a quarter-wave plate placed in front of them.}
\label{fig:Sketch}
\end{figure}

Figure~\ref{fig:Sketch} shows a schematic of the realized amplifier. The input beam is linearly polarized, has a beam quality of $\text{M}^2 < 1.1$ and a 1/$\text{e}^2$ radius of 2.75\,mm \cite{Zeyen2023a}. It is injected into the amplifier through a thin-film polarizer (TFP) and relay imaged to the disk via the 4f-stage (consisting of $\text{M}_\text{cav, 1}$ and $\text{M}_\text{cav, 2}$ separated by two plane folding mirrors $\text{M}_\text{plan, 1}$ and $\text{M}_\text{plan, 1}$). The beam then enters the FT part of the amplifier where it is guided from the disk, via the convex mirror $\text{M}_\text{vex}$ and a folding mirror, to array mirror 1. From there, the beam is directed to end-mirror $\text{M}_{2,1}$ which is equipped with a quarter-wave plate that ensures 90° rotation of the polarization after the double pass. Upon reflection again at the disk, the beam has undergone a Fourier transform (i.e., the profile of the beam leaving the disk corresponds to the FT of the beam impinging for the first time on the disk). Since the polarization was rotated, the beam is now imaged to end-mirror $\text{M}_1$ undergoing a reflection at the TFP. In the following round trips, the 4f-stage allows angular multiplexing on the disk. Thus, a second mirror array in the 4f branch of the amplifier is not required. The small angular spread of the beams on the disk is amplified by $\text{M}_\text{vex}$. The mirror array sends the beam alternately to $\text{M}_{2,2}$ and $\text{M}_{2,3}$ and the beam is kept circulating in the amplifier by reflecting at the TFP. On the final pass, the beam is sent over $\text{M}_{2,4}$, which is also equipped with a quarter-wave plate. In a double-pass through this wave plate, the polarization of the beam is rotated back so that the beam is transmitted by the TFP and exits the amplifier. Since $\text{M}_{2,1}$ and $\text{M}_{2,4}$ are separated vertically, the in- and out-going beams propagate under different angles and can easily be separated. The propagation scheme over the mirror array and the end-mirrors is sketched in the inset of  Fig.~\ref{fig:Sketch}.

While purely FT based amplifier designs are well known to deliver high-quality beams \cite{Schuhmann_thin-disk_2015, Schuhmann:19}, the addition of a 4f-stage to such a multipass amplifier is new and provides several advantages. For example, as mentioned, our multipass amplifier design requires on average only one array mirror per pass over the disk thanks to the 4f-stage. This is half the number of mirrors used in previous resonator-based designs \cite{antognini_thin-disk_2009, schuhmann_thin-disk_2017, Schuhmann_thin-disk_2015}. Furthermore, the 4f-stage allows for small incidence angles on the disk so that the angular spread between individual passes is small. The mirror array can thus be placed relatively far from the disk, in the telescope part of the resonator segment where the beam diameter is small, and 0.5'' diameter array mirrors can be used. This minimizes astigmatism and reduces both cost and footprint of the multipass amplifier.

\begin{figure}[h!]
\centering\includegraphics[width=\linewidth]{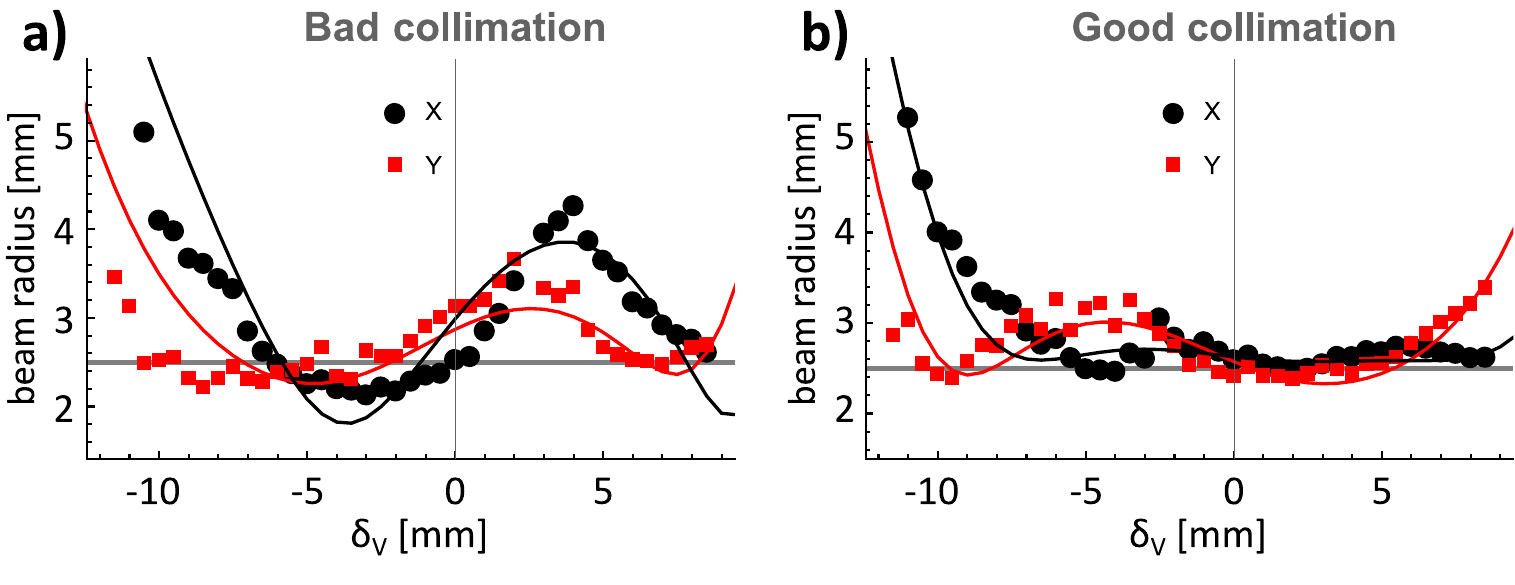}
\caption{Beam size at the disk on the $6^\text{th}$ pass vs. effective focal power of the disk for the two transverse directions (X, Y). The effective dioptric power of the disk was scanned by varying $\delta_\text{V}$, the separation between the imaging mirrors of the 4f-stage. The solid lines are fits with input beam diameter $w_0 = 5.5$\,mm and beam divergence a) $\theta = 0.15$\,mrad and b) $\theta = 0.126$\,mrad (note that a “collimated” Gaussian beam has $\theta = \lambda / \pi w_0 \neq 0$).}
\label{fig:Collimation}
\end{figure}

The 4f-stage also helps to mode match the input beam to the eigenmode of the multipass amplifier. The key idea is to measure the beam size on the unpumped disk versus effective focal power of the disk~\cite{zeyenthin}, where the distance between the focusing elements in the 4f-part is varied to tune $\Delta V$ according to~\eqref{eq:det_4f} (see Figure~\ref{fig:Collimation}). In such plots, the beam size of a properly mode matched beam will show minimal oscillation around $\Delta V = 0$, i.e., around the nominal value of the focal length of the disk. Using this procedure, the disk cannot be pumped, since the soft aperture of the pump spot would suppress these oscillations \cite{schuhmann_multipass_2018, zeyenthin}. In this data set, the beam size was measured after 6~passes over the disk as a compromise between sensitivity (sufficiently large oscillation signal) and time needed for alignment. In Fig.~\ref{fig:Collimation}a, the input beam is estimated to converge with half angle $\theta = 0.15$\,mrad, which explains the relatively large oscillations in beam size \cite{zeyenthin}. In Fig.~\ref{fig:Collimation}b, the collimation of the input beam is optimized by slightly moving one of the telescope lenses used to mode match the input beam to the amplifier. The oscillations around $\delta_\text{V} = 0$\,mm are suppressed, meaning that the beam is correctly injected. The beam size is different in X (horizontal) and Y (vertical) directions because of a slight astigmatism, mainly introduced by the 4f-stage.

\section{Experimental results} \label{sec:Results}

\begin{figure}[h!]
\centering\includegraphics[width=0.6\linewidth]{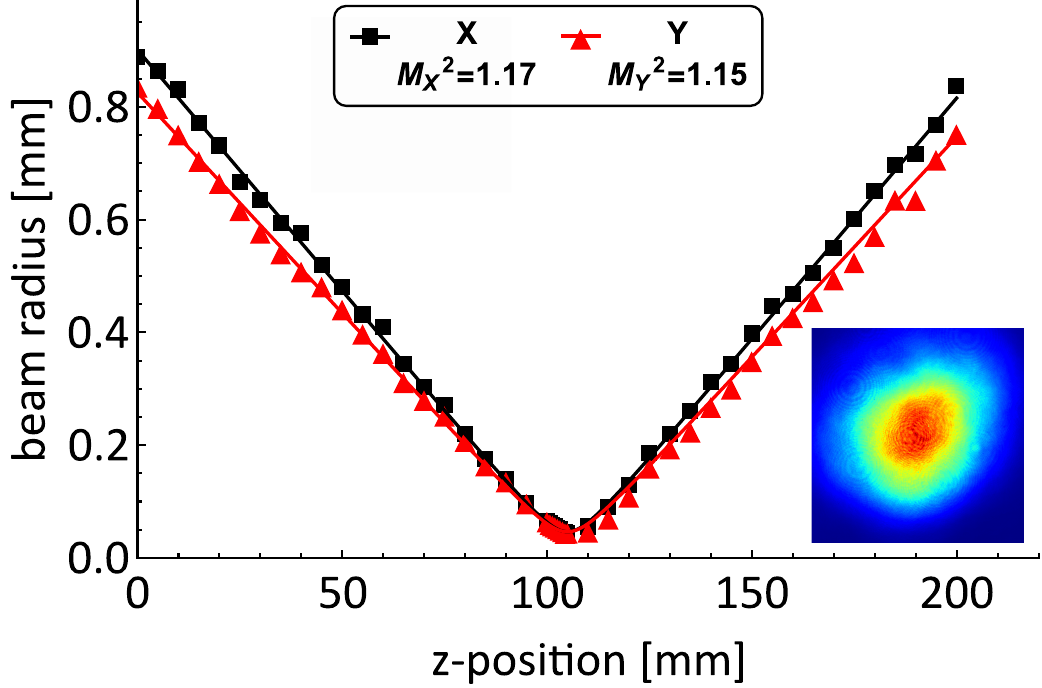}
\caption{$\text{M}^2$ measurement of the output beam of the 20-pass hybrid-type amplifier pumped at 3.5\,kW/$\text{cm}^2$. Inset: Far-field beam profile with 1/$\text{e}^2$ radius of 2.75\,mm.}
\label{fig:MSquared}
\end{figure}

With a properly mode matched input beam with a waist of $w_0 = 2.75$\,mm at the disk, a small signal gain of about 20 was reached with 20 passes (reflections) over the disk. As shown in Fig.~\ref{fig:MSquared}, the output beam was close to diffraction limited with $\text{M}^2 \leq 1.17$ in horizontal and vertical direction (measurement taken according to ISO 11146 with Thorlabs BP209-IR2). The pump radiation is delivered by a volume Bragg grating stabilized diode stack running at 969\,nm at the zero-phonon line of Yb:YAG. Compared to conventional pumping at 940\,nm, the heat load is reduced by about 30\,\%, which further reduces thermal lensing effects in the amplifier. Still the pump intensity was limited to 3.5\,kW/$\text{cm}^2$ (pump spot diameter of 8\,mm at full width half maximum) to avoid overheating the water-cooled disk.

\begin{figure}[h!]
\centering\includegraphics[width=0.7\linewidth]{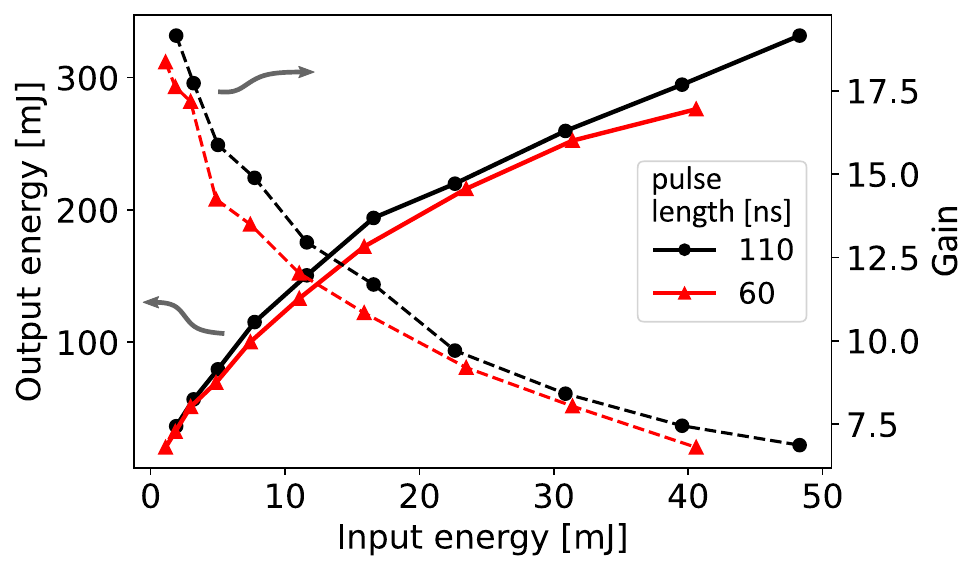}
\caption{Output pulse energy and gain of the hybrid 20-pass amplifier versus input pulse energy for 110\,ns and 60\,ns long input pulses. A pulsed gain of 6.9 was achieved for 110\,ns pulses at a output pulse energy of 330\,mJ. At around 275\,mJ, optically induced breakdown of the air in the focus of the 4f-stage started to take place. The disk was 600\,\textmu m thick with 2.3\,\% doping, the beam size on the disk was $w_0 = 3$\,mm and the pump intensity was 2.5\,kW/$\text{cm}^2$ at 1300\,W pump power.}
\label{fig:Energy}
\end{figure}

Seeded by pulses from our single-frequency TDL \cite{Zeyen2023a}, the hybrid 20-pass amplifier delivered single frequency pulses of 60\,ns and 110\,ns length at 275\,mJ and 330\,mJ, respectively. The repetition rate was fixed to 100~Hz. To the best of our knowledge, this is the highest single frequency pulse energy obtained from a TDL to date. The pulsed gain was about 7 at 40\,mJ input pulse energy for both pulse lengths. Figure~\ref{fig:Energy} shows how the gain (red) decreases as the extracted pulse energy (black) increasingly depletes the inversion in the disk. At around 275\,mJ, the 60\,ns pulses ionized the air in the focus of the 4f-stage, which prevented further energy scaling. Assuming a diffraction limited spot size and basic scaling laws for laser induced breakdown of air \cite{Niemz:95} this energy approximately agrees with expectations. The input pulse energy was not increased above 50\,mJ to avoid laser-induced damage in the oscillator.

\begin{figure}[h!]
\centering\includegraphics[width=0.8\linewidth]{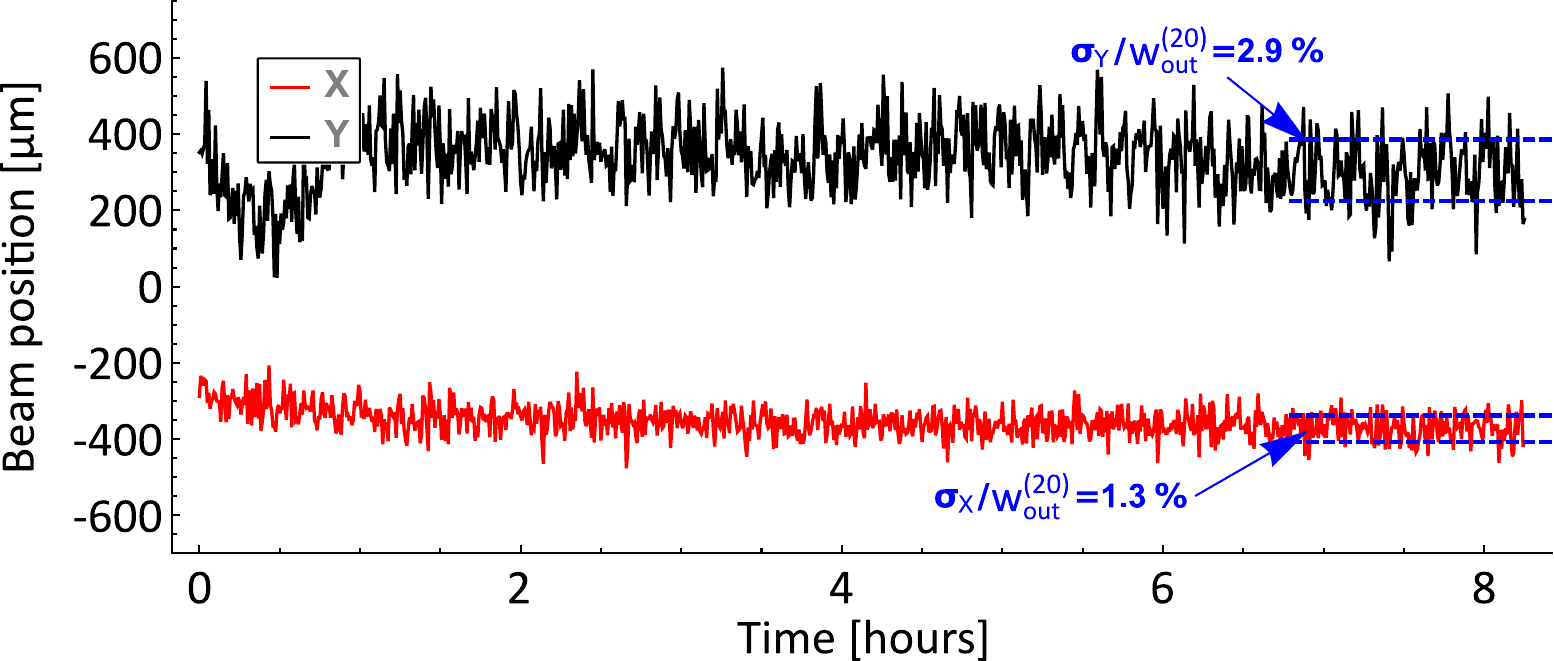}
\caption{Measured output beam position after 20 passes under full pump load and low-power CW operation. The jitter is about 2.9\,\% of the beam radius (2.75\,mm) in the vertical (Y-direction) and about 1.3\,\% in the horizontal (X-direction).}
\label{fig:Pointing}
\end{figure}

After aligning the amplifier under full pump load, the system was continuously operated overnight, and showed only marginal misalignment over this period, as can be seen in Fig.~\ref{fig:Pointing}. The pointing jitter was only $<3$\,\% and $< 1.5$\,\% of the beam diameter (in this case 5.5 mm) in the vertical and horizontal direction, respectively, underlining the good stability against small misalignments of this amplifier. Thanks to the compact mirror array and the space saving 4f-propagation, the footprint of the multipass amplifier was about $1\,\text{m} \times 0.4\,\text{m}$, including the pump optics.

\section{Discussion and conclusion} \label{sec:Conclusion}

We demonstrated, for the first time, a compact hybrid multipass amplifier based on a succession of optical Fourier transform and 4f-relay imaging. While we realized a thin-disk amplifier with this architecture, the general scheme should be applicable to most laser types. Fourier transforming the beam in one part of the amplifier allows a passive compensation of phase front curvature errors and makes this amplifier suited for high-power applications where excellent beam quality is required. The 4f-stage in the other part of the amplifier is used to compensate shifts in focal length of the pumped disk and helps to mode match the injected beam. It also helps keeping the whole system compact. Moreover, changing the effective focal length of the disk by detuning the 4f-stage makes the system more flexible. If needed, $f_\text{eff}$ and the resonator layout can be adjusted to increase the beam size on the end mirrors in the FT part of the amplifier without changing the optics, to avoid laser-induced damage.

\section*{Funding}

The European Research Council (ERC) through CoG. \#725039; the Swiss National Science Foundation through the projects SNF 200021\_165854 and SNF 200020\_197052; the Deutsche Forschungsgemeinschaft (DFG, German Research Foundation) under Germany's Excellence Initiative EXC 1098 PRISMA (194673446), Excellence Strategy EXC PRISMA+ (390831469) and DFG/ANR Project LASIMUS (DFG Grant Agreement 407008443); the French National Research Agency with project ANR-18-CE92-0030-02;

\section*{Acknowledgments}
We gratefully acknowledge the support of the ETH Zürich electronics workshop. In particular, we thank Diogo Di Calafiori and Dr. Werner Lustermann. We also thank Dr. Marcos Gaspar, Dr. Carlo Vicario, Dr. Cezary Sidle and Stefan Mair from PSI.

\section*{Disclosures}

The authors declare that there are no conflicts of interest related to this article.

\section*{Data Availability Statement}

The data that support the findings of this study are available from the corresponding author upon reasonable request.



\begin{thebibliography}{10}
\newcommand{\enquote}[1]{``#1''}

\bibitem{Zapata:15}
L.~E. Zapata, H.~Lin, A.-L. Calendron, H.~Cankaya, M.~Hemmer, F.~Reichert,
  W.~R. Huang, E.~Granados, K.-H. Hong, and F.~X. K\"{a}rtner,
  \enquote{Cryogenic {Y}b:{YAG} composite-thin-disk for high energy and average
  power amplifiers,} {\protect\JournalTitle{Opt. Lett.}} \textbf{40},
  2610--2613 (2015).

\bibitem{Herkommer:20}
C.~Herkommer, P.~Kr\"{o}tz, R.~Jung, S.~Klingebiel, C.~Wandt, R.~Bessing,
  P.~Walch, T.~Produit, K.~Michel, D.~Bauer, R.~Kienberger, and T.~Metzger,
  \enquote{Ultrafast thin-disk multipass amplifier with 720 m{J} operating at
  kilohertz repetition rate for applications in atmospheric research,}
  {\protect\JournalTitle{Opt. Express}} \textbf{28}, 30164--30173 (2020).

\bibitem{Nagel:21}
S.~Nagel, B.~Metzger, D.~Bauer, J.~Dominik, T.~Gottwald, V.~Kuhn, A.~Killi,
  T.~Dekorsy, and S.-S. Schad, \enquote{Thin-disk laser system operating above
  10 k{W} at near fundamental mode beam quality,} {\protect\JournalTitle{Opt.
  Lett.}} \textbf{46}, 965--968 (2021).

\bibitem{siebold_high-energy_2012}
M.~Siebold, M.~Loeser, F.~Roeser, M.~Seltmann, G.~Harzendorf, I.~Tsybin,
  S.~Linke, S.~Banerjee, P.~D. Mason, P.~J. Phillips, K.~Ertel, J.~C. Collier,
  and U.~Schramm, \enquote{High-energy, ceramic-disk {Yb}:{LuAG} laser
  amplifier,} {\protect\JournalTitle{Optics Express}} \textbf{20}, 21992
  (2012).

\bibitem{ochi_effective_2017}
Y.~Ochi, K.~Nagashima, M.~Maruyama, and R.~Itakura, \enquote{Effective
  {Multi}-pass {Amplification} {System} for {Yb}:{YAG} {Thin}-{Disk} {Laser},}
  in \emph{Laser {Congress} 2017 ({ASSL}, {LAC}),}  (OSA, Nagoya, Aichi, 2017),
  p. JTh2A.31.

\bibitem{perevezentsev_matrix_2017}
E.~Perevezentsev, I.~Kuznetsov, I.~Mukhin, and O.~V. Palashov, \enquote{Matrix
  multi-pass scheme disk amplifier,} {\protect\JournalTitle{Applied Optics}}
  \textbf{56}, 8471 (2017).

\bibitem{papadopoulos_high_2015}
D.~N. Papadopoulos, F.~Friebel, A.~Pellegrina, M.~Hanna, P.~Camy, J.-L.
  Doualan, R.~Moncorge, P.~Georges, and F.~P. H.~J. Druon, \enquote{High
  {Repetition} {Rate} {Yb}:{CaF}$_{\textrm{2}}$ {Multipass} {Amplifiers}
  {Operating} in the 100-{mJ} {Range},} {\protect\JournalTitle{IEEE Journal of
  Selected Topics in Quantum Electronics}} \textbf{21}, 464--474 (2015).

\bibitem{Zwilich:20}
M.~Zwilich and B.~Ewers, \enquote{Coherent beam combining of multipass
  thin-disk lasers with active phase control,} {\protect\JournalTitle{OSA
  Continuum}} \textbf{3}, 3176--3186 (2020).

\bibitem{druon_comparison_2020}
F.~Druon, K.~Genevrier, P.~Georges, and D.~N. Papadopoulos, \enquote{Comparison
  of multi-pass and regenerative strategies for energetic high-gain amplifiers
  based on {Yb}:{CaF}$_{\textrm{2}}$,} {\protect\JournalTitle{Opt. Lett.}}
  \textbf{45}, 4408 (2020).

\bibitem{nagel_thin_2019}
S.~Nagel, B.~Metzger, T.~Gottwald, V.~Kuhn, A.~Killi, and S.-S. Schad,
  \enquote{Thin disk laser operating in fundamental mode up to a power of 4
  k{W},} in \emph{The European Conference on Lasers and Electro-Optics,}
  (Optical Society of America, 2019), p. ca\_5\_4.

\bibitem{negel_ultrafast_2015}
J.-P. Negel, A.~Loescher, A.~Voss, D.~Bauer, D.~Sutter, A.~Killi, M.~A. Ahmed,
  and T.~Graf, \enquote{Ultrafast thin-disk multipass laser amplifier
  delivering 1.4 {kW} (47 {mJ}, 1030 nm) average power converted to 820 {W} at
  515 nm and 234 {W} at 343 nm,} {\protect\JournalTitle{Optics Express}}
  \textbf{23}, 21064 (2015).

\bibitem{schulz_pulsed_2012}
M.~Schulz, R.~Riedel, A.~Willner, S.~D{\"u}sterer, M.~J. Prandolini,
  J.~Feldhaus, B.~Faatz, J.~Rossbach, M.~Drescher, and F.~Tavella,
  \enquote{Pulsed operation of a high average power {Yb}:{YAG} thin-disk
  multipass amplifier,} {\protect\JournalTitle{Optics Express}} \textbf{20},
  5038 (2012).

\bibitem{dietz_ultrafast_2020}
T.~Dietz, M.~Jenne, D.~Bauer, M.~Scharun, D.~Sutter, and A.~Killi,
  \enquote{Ultrafast thin-disk multi-pass amplifier system providing 19 {kW} of
  average output power and pulse energies in the 10 {mJ} range at 1 ps of pulse
  duration for glass-cleaving applications,} {\protect\JournalTitle{Opt.
  Express}} \textbf{28}, 11415 (2020).

\bibitem{hornung_54_2016}
M.~Hornung, H.~Liebetrau, S.~Keppler, A.~Kessler, M.~Hellwing, F.~Schorcht,
  G.~A. Becker, M.~Reuter, J.~Polz, J.~K{\"o}rner, J.~Hein, and M.~C. Kaluza,
  \enquote{54 {J} pulses with 18 nm bandwidth from a diode-pumped chirped-pulse
  amplification laser system,} {\protect\JournalTitle{Opt. Lett.}} \textbf{41},
  5413 (2016).

\bibitem{antognini_thin-disk_2009}
A.~Antognini, K.~Schuhmann, F.~D. Amaro, F.~Biraben, A.~Dax, A.~Giesen,
  T.~Graf, T.~W. Hansch, P.~Indelicato, L.~Julien, C.-Y. Kao, P.~E. Knowles,
  F.~Kottmann, E.~Le~Bigot, Y.-W. Liu, L.~Ludhova, N.~Moschuring, F.~Mulhauser,
  T.~Nebel, F.~Nez, P.~Rabinowitz, C.~Schwob, D.~Taqqu, and R.~Pohl,
  \enquote{Thin-{Disk} {Yb}:{YAG} {Oscillator}-{Amplifier} {Laser}, {ASE}, and
  {Effective} {Yb}:{YAG} {Lifetime},} {\protect\JournalTitle{IEEE Journal of
  Quantum Electronics}} \textbf{45}, 993--1005 (2009).

\bibitem{Schuhmann_thin-disk_2015}
K.~Schuhmann, M.~Ahmed, A.~Antognini, T.~Graf, T.~H{\"a}nsch, K.~Kirch,
  F.~Kottmann, R.~Pohl, D.~Taqqu, A.~Voss \emph{et~al.}, \enquote{Thin-disk
  laser multi-pass amplifier,} in \emph{Solid State Lasers XXIV: Technology and
  Devices,}  vol. 9342 (International Society for Optics and Photonics, 2015),
  p. 93420U.

\bibitem{chyla_generation_2018}
M.~Chyla, S.~S. Nagisetty, P.~Severov{\'a}, H.~Zhou, M.~Smr{\v z}, A.~Endo, and
  T.~Mocek, \enquote{Generation of 1-{J} bursts with picosecond pulses from
  {Perla} {B} thin-disk laser system,} in \emph{Solid {State} {Lasers} {XXVII}:
  {Technology} and {Devices},}  W.~A. Clarkson and R.~K. Shori, eds. (SPIE, San
  Francisco, United States, 2018), p.~21.

\bibitem{Loescher:22}
A.~Loescher, F.~Bienert, C.~R\"{o}cker, T.~Graf, M.~Gorjan, J.~A. der Au, and
  M.~A. Ahmed, \enquote{Thin-disk multipass amplifier delivering sub-400 fs
  pulses with excellent beam quality at an average power of 1 k{W},}
  {\protect\JournalTitle{Opt. Continuum}} \textbf{1}, 747--758 (2022).

\bibitem{schuhmann_multipass_2018}
K.~Schuhmann, K.~Kirch, M.~Marszalek, F.~Nez, R.~Pohl, I.~Schulthess, L.~{\v
  S}ink{\= u}nait{\. e}, G.~Wichmann, M.~Zeyen, and A.~Antognini,
  \enquote{Multipass amplifiers with self-compensation of the thermal lens,}
  {\protect\JournalTitle{Applied Optics}} \textbf{57}, 10323 (2018).

\bibitem{zeyenthin}
M.~Zeyen, \enquote{Thin-disk laser for the measurement of the
  hyperfine-splitting in muonic hydrogen,} Ph.D. thesis, ETH Zurich (2021).

\bibitem{amaro_laser_2022}
P.~Amaro, A.~Adamczak, M.~A. Ahmed, L.~Affolter, F.~D. Amaro, P.~Carvalho,
  T.~L. Chen, L.~M.~P. Fernandes, M.~Ferro, D.~Goeldi, T.~Graf, M.~Guerra,
  T.~W. Hänsch, C.~A.~O. Henriques, Y.~C. Huang, P.~Indelicato, O.~Kara,
  K.~Kirch, A.~Knecht, F.~Kottmann, Y.~W. Liu, J.~Machado, M.~Marszalek,
  R.~D.~P. Mano, C.~M.~B. Monteiro, F.~Nez, J.~Nuber, A.~Ouf, N.~Paul, R.~Pohl,
  E.~Rapisarda, J.~M.~F. dos Santos, J.~P. Santos, P.~A. O.~C. Silva,
  L.~Sinkunaite, J.~T. Shy, K.~Schuhmann, S.~Rajamohanan, A.~Soter, L.~Sustelo,
  D.~Taqqu, L.~B. Wang, F.~Wauters, P.~Yzombard, M.~Zeyen, and A.~Antognini,
  \enquote{{Laser excitation of the 1s-hyperfine transition in muonic
  hydrogen},} {\protect\JournalTitle{SciPost Phys.}} \textbf{13}, 020 (2022).

\bibitem{nuber2022diffusion}
J.~Nuber, A.~Adamczak, M.~A. Ahmed, L.~Affolter, F.~D. Amaro, P.~Amaro,
  A.~Antognini, P.~Carvalho, Y.~H. Chang, T.~L. Chen, W.~L. Chen, L.~M.~P.
  Fernandes, M.~Ferro, D.~Goeldi, T.~Graf, M.~Guerra, T.~W. Hänsch, C.~A.~O.
  Henriques, M.~Hildebrandt, P.~Indelicato, O.~Kara, K.~Kirch, A.~Knecht,
  F.~Kottmann, Y.~W. Liu, J.~Machado, M.~Marszalek, R.~D.~P. Mano, C.~M.~B.
  Monteiro, F.~Nez, A.~Ouf, N.~Paul, R.~Pohl, E.~Rapisarda, J.~M.~F. dos
  Santos, J.~P. Santos, P.~A. O.~C. Silva, L.~Sinkunaite, J.~T. Shy,
  K.~Schuhmann, S.~Rajamohanan, A.~Soter, L.~Sustelo, D.~Taqqu, L.~B. Wang,
  F.~Wauters, P.~Yzombard, M.~Zeyen, and J.~Zhang, \enquote{{Diffusion of
  muonic hydrogen in hydrogen gas and the measurement of the 1$s$ hyperfine
  splitting of muonic hydrogen},} {\protect\JournalTitle{SciPost Phys. Core}}
  \textbf{6}, 057 (2023).

\bibitem{Zeyen2023}
M.~Zeyen, L.~Affolter, M.~A. Ahmed, T.~Graf, O.~Kara, K.~Kirch, M.~Marszalek,
  F.~Nez, A.~Ouf, R.~Pohl, S.~Rajamohanan, P.~Yzombard, A.~Antognini, and
  K.~Schuhmann, \enquote{{Pound–Drever–Hall locking scheme free from Trojan
  operating points},} {\protect\JournalTitle{Review of Scientific Instruments}}
  \textbf{94}, 013001 (2023).

\bibitem{Zeyen2023a}
M.~Zeyen, L.~Affolter, M.~A. Ahmed, T.~Graf, O.~Kara, K.~Kirch, A.~Langenbach,
  M.~Marszalek, F.~Nez, A.~Ouf, R.~Pohl, S.~Rajamohanan, P.~Yzombard,
  K.~Schuhmann, and A.~Antognini, \enquote{Injection-seeded high-power
  {Y}b:{YAG} thin-disk laser stabilized by the {P}ound-{D}rever-{H}all method,}
  {\protect\JournalTitle{Opt. Express}} \textbf{31}, 29558--29572 (2023).

\bibitem{schuhmann_thin-disk_2017}
K.~Schuhmann, \enquote{The {Thin}-{Disk} {Laser} for the {2S} {\textendash}
  {2P} {Measurement} in {Muonic} {Helium},} Ph.D. thesis, ETH Zurich (2017).

\bibitem{neuhaus_passively_2009}
J.~Neuhaus, \enquote{Passively mode-locked {Yb}:{YAG} thin-disk laser with
  active multipass geometry,} {PhD} {Thesis}, Universit{\"a}t Konstanz,
  Konstanz (2009).

\bibitem{Schuhmann:19}
K.~Schuhmann, K.~Kirch, A.~Knecht, M.~Marszalek, F.~Nez, J.~Nuber, R.~Pohl,
  I.~Schulthess, L.~Sinkunaite, M.~Zeyen, and A.~Antognini, \enquote{Passive
  alignment stability and auto-alignment of multipass amplifiers based on
  {F}ourier transforms,} {\protect\JournalTitle{Appl. Opt.}} \textbf{58},
  2904--2912 (2019).

\bibitem{Niemz:95}
M.~H. Niemz, \enquote{{Threshold dependence of laser‐induced optical
  breakdown on pulse duration},} {\protect\JournalTitle{Applied Physics
  Letters}} \textbf{66}, 1181--1183 (1995).

\end{thebibliography}
\end{document}